\documentstyle[12pt,epsfig,fleqn]{article} \oddsidemargin .4cm \topmargin
         0cm \headsep .5cm  \textheight 21.5cm \textwidth 14.9cm
   \newcommand{\be}{\begin{eqnarray}}  \newcommand{\ee}{\end{eqnarray}}
 \newcommand{\nee}{\nonumber\end{eqnarray}} \newcommand{\nn}{\nonumber\\}
             \newcommand\sfrac[2]{{\textstyle \frac{#1}{#2}}}
    \newcommand{\dgs}{d^{\gamma}{\!\scriptstyle (}s{\scriptstyle )}\,}
      \newcommand{\dzs}{d^{Z}{\!\scriptstyle (}s{\scriptstyle )}\,}
   \newcommand{\dgz}{d^{\gamma,Z}{\!\scriptstyle (}s{\scriptstyle )}\,}
               \newcommand{\mIm}{\,\mbox{\small $\Im$m\,}}
               \newcommand{\eRe}{\,\mbox{\small $\Re$e\,}}
   
   \newcommand{\Tr}{\mbox{\small Tr}} \newcommand{\CP}{$CP$ violation}
 \newcommand{\plmin} {\mbox{ \mbox{\raisebox{-0.1ex}{${\scriptscriptstyle
       (}\!$}} \mbox{\raisebox{ 0.2ex}{$\!{\displaystyle \pm}\!$}}
         \mbox{\raisebox{-0.1ex}{$\!{\scriptscriptstyle )}$}} }}

\begin{document}

\begin{titlepage}
\begin{flushright}INRNE-TH-98/33\\
September 1998\\
hep-ph/9809290
\end{flushright}\vfill

\vspace{2cm}
\begin{center}
{\Large\bf CP violating asymmetries  of  $b$ quarks and leptons\\ 
 in $e^+e^- \rightarrow t\bar t$  and supersymmetry}

\vspace{1cm}
{\em\Large review}
\vspace{2.5cm}

{\large Ekaterina Christova}\\ 
\vspace{0.5cm}
{\em Inst. of Nucl. Res. and Nucl. Energy, Bulg. Academy of Sciencies,\\ 
      Tzarigradsko chaussee 72, Sofia 1784, Bulgaria
\footnote{permanent address}\\
and\\
International Center for Theoretical Physics, 
  Trieste, Ttaly}

\vfill

{\sc Abstract}\end{center}
The distributions of the single decay b-quarks and leptons from 
$e^+e^- \to t\bar t$
assuming CP violation are reviewed. Different asymmetries , sensitive 
independently to CP violation in the production and in the decay,
 and sensitive to the real and imaginary parts of 
$d^\gamma$ and $d^Z$ are defined.
The analytic expressions  
 are general and independent on the model of CP violation. 
In most of them all phase space integrations are
 fulfilled analytically. Numerical results in the 
MSSM with complex couplings are presented. 

\end{titlepage}
\newpage\setcounter{footnote}{0}\setcounter{page}{1}
                                          \setcounter{page}{1}

\tableofcontents
\newpage


\section{Introduction} 
  The large mass of the top quark allows to probe high energies, where
      new physics might show up as well. In the last years a number of
  papers consider CP violating observables in processes with top quarks
 as tests of physics beyond the Standard Model (SM).
  There are several reasons for this:

\begin{enumerate}

 \item 
Owing to its large mass ($m_t= 175 GeV$) \cite{mtop},
 the top quark will decay 
before forming a hadronic bound state. Therefore its polarization will
      not be deluted by possible hadronization processes and it can be
   determined by the distributions of its decay products. CP violation
                    is sensitive to the polarization of the top-quark.
 \item 
Theoretical predictions are more reliable as they are free of
   the hadronization uncertainties.
 
\item
  Due to the GIM mechanism, the effects of CP violation in SM are very
 small.  Thus, observation of CP violation  in the top-quark
         physics would be an indication of physics beyond the
 SM. Supersymmetric models and models with more than one Higgs doublet
        are at present the most favoured candidates. They  provide new
  sources of CP violation \cite{Dugan} that lead to CP non conservation
 at one loop    level.

\end{enumerate}

For testing  of CP invariance one has to compare the decays of the top
  quark with those of the anti-top quark.  It is important that in the
      future $e^+e^-$ colliders   the top--antitop quark pairs will be
 copiously produced and their decay modes  will be studied in the same
                                                           experiment.

      In general CP violation  enters both  the production and decay
   processes. In  the distribution of the $t$-decay products it enters
  through the $t$-polarization. This means that if the $t$-quark would
decay unpolarized (due to  possible hadronization processes) 
the distribution of its decay
             products would not be sensitive to possible CP violation. 
As shown in~\cite{Bigi} the effects of depolarization due to 
 such hadronization 
processes for the top quark  are very small.

    We shall consider the $t$ ($\bar t$) - quarks produced in $e^+e^-$
annihilation\footnote{CP violation in $t\bar t$ production at hadron
 colliders has been discussed in  \cite{hadrons}}. 
They will be identified by their decay products.  As the
        top quark actually does not mix with the quarks from the other
   generations, its only decay mode in SM is $t\to bW$.  Consequently,
   information about the $t$-polarization  will be carried both by the
    $b$-quarks  and by the $W$'s.  We shall consider the general
 expressions for  CP
 violation comparing  the angular~\cite{angle} 
and energy~\cite{energy}  distributions of $b$ and
     $\bar b$ in the CP conjugate processes: 
\be
 e^+ + e^- \rightarrow
t{\bar t} \rightarrow b + X'\,,\qquad  e^+ + e^- \rightarrow t{\bar t}
     \rightarrow {\bar b} + {\bar X'}\,.\label{b}
 \ee 
$X'$ ($\bar X'$)
      stands for $\bar t W$  ($tW^-$) irrespectively how the $W's$ are
   identified. Previously the effects of the 
dipole moment form factors in (\ref{b}) 
 were considered in \cite{Soni}.  
 The CP violating asymmetries organized from 
 the $b$ and $\bar b$ quarks in processes (\ref{b}) have the advantage
that they have 
no background: If the $b$ and $\bar b$ quarks are produced directly in
 $e^+e^-\to b\bar b$ the final state is CP even and thus cannot induce
 any CP violating asymmetries. The $b$ and $\bar b$ quarks may come 
also from the decays  of $W^\pm$: $e^+e^- \to W^+W^-$, $W^\pm \to bc$,
 that are  Cabbibo supressed and CP violation can be of academic
 interest only. Thus, measuring CP violation through   
(\ref{b}) does not require reconstruction of the processes event by event. 
 However, a clear identification of the jets from $b$ and $\bar b$ is
         necessary. Different methods of $b$-tagging are considered in 
\cite{angle}.

The $W$'s can be  studied through the angular 
\cite{{Bern},{we},{many},{Rindani}} and energy \cite{{Chang},{Sehgal-2},
{Grz-E}}   distributions of the leptons from
        the decays $W^\pm\to l^\pm \nu$:
\be 
 e^+ + e^- \rightarrow t\bar t \rightarrow bl^+X\,,
      \qquad  e^+ + e^- \rightarrow t\bar t \rightarrow 
\bar b l^-\bar  X\label{l}
 \ee 
  Here we shall  study CP violation
      comparing the angular distributions of $b, l^+$ and $\bar b,l^-$
  with special emphasis on triple product correlations \cite{we}.
 
 The longitudinal polarizations of $e^+$ and  $e^-$ are also taken 
into account.

  There are three verticies that can introduce \CP:
\begin{itemize}
\item
      In  the $\gamma t\bar t$ and $Zt\bar t$ vertices CP violation is
introduced by the electric $\dgs$ and weak $\dzs$   dipole moment form
 factors:
 \be  
 e \,  {\cal
 V}^{\gamma}_{\mu}   &=& e \,  \left( \frac{2}{3} \gamma_{\mu}  - i \,
            \frac{\dgs}{m_{t}} {\cal P}_{\mu} \gamma_{5}  \right) \, ,
      \label{vertexgamma} \\  
 g_Z \,  {\cal V}^{Z}_{\mu}   &=&g_Z \,
            \left(  \gamma_{\mu} ( g_{V} + g_{A} \gamma_{5} )   - i \,
             \frac{\dzs}{m_{t}} {\cal P}_{\mu} \gamma_{5}   \right) \,
          ,\label{vertexZ} 
 \ee 
 where {${\cal P}_{\mu} = p_{t\,\mu} -
   p_{\bar{t}\,\mu}$},   {$g_{V} = (1/2) - (4/3) \sin^{2}\Theta_{W}$},
    {$g_{A} = - (1/2)$}, and  {$g_Z = e/\sin2\Theta_{W}$} with $e$ the
    electro--magnetic  coupling constant. The electoweak dipole moment
  form factors  $\dgz$ are functions of $s$, so that $d^\gamma (0)$ and
   $d^Z (m_Z^2)$ determine the electric and weak dipole moments of the
top quark. $\dgz$ can be introduced only by  an  interaction
                       in the production process $e^+e^- \to t\bar t$
 that is both $P$ and $T$ violating, and through $CPT$
 invariance, also $CP$ violating. Very recently a nice review
 on the dipole moment form factors appeared~\cite{Hollik}.

\item 
 The $tbW$ vertex, that determine the weak decays of the $t$ and $\bar
               t$ quarks, see eq. (\ref{Mtop}),
 is  written in the form: 
\be 
V^{t}_\alpha & =&
                 \frac{g}{2\sqrt{2}}\left(\gamma_\alpha (1-\gamma_5) +
   f_L^t\gamma_\alpha (1-\gamma_5 ) + \frac{g_R^{t}}{m_W} P^{t}_\alpha
            (1+ \gamma_5 )\right) ,\label{t} 
\\ V^{\bar{t}}_\alpha &=&
                 \frac{g}{2\sqrt{2}}\left(\gamma_\alpha (1+\gamma_5) +
f_L^{\bar{t}*}\gamma_\alpha (1+\gamma_5 ) + \frac{g_R^{\bar{t}*}}{m_W}
                             {P}^{\bar{t}}_\alpha (1- \gamma_5)\right)
\label{tbar}
                         \ee 
with  ${P}^t = p_t+p_b$, ${P}^{\bar{t}} =
 p_{\bar{t}}+p_{\bar{b}}$.  In eqs.~(\ref{t}) and (\ref{tbar}) we have
    kept only the terms that do not vanish in the approximation $m_b =
    0$. Contrary to the electroweak dipole moment form factors $\dgz$,
   the form factors $f_L^{t,\bar{t}} $ and $g_R^{t,\bar{t}}$ have both
$CP$--invariant and $CP$--violating contributions: 
\be  
f_L^{t,\bar t}
        = f_L^{SM} \pm if_L^{CP}\,\qquad g_R^{t,\bar t} = g_R^{SM} \pm
   ig_R^{CP}\, 
\ee 
where the superscript {\scriptsize SM (CP)} denotes
  the CP invariant (CP violating) contributions to the form factors. In
      analogy with the dipole moment form factors, we have  explicitely
  written the i in front of $f_L^{CP}$ and $g_R^{CP}$, that comes from
   the imaginary CP violating coupling.   Both the CP invariant and CP
 violating parts of the fromfactors  $f_L^{SM(CP)}$ and $g_R^{SM(CP)}$
  have real and imaginary parts. If neglecting absorptive parts of the
 amplitude,  $\Im m f_L^{SM} = \Im m f_L^{CP} =0 $ and $\Im m g_R^{SM}
          = \Im m g_R^{CP} =0 $,  then CP invariance  implies that the
                 form factors of the top and anitop quarks are real and
          equal: 
 \be
 f_L^t=f_L^{\bar{t}} = \Re e
            f_L^{SM},\qquad g_R^t=g_R^{\bar{t}} = \Re e\, g_R^{SM} 
\ee
\end{itemize}

   For understanding the mechanism of CP violation, it is important to
  distinguish CP violation in the production from CP violation  in the
      decay processes. As we shall see, the distributions of the decay
  products are mainly sensitive to CP violation in the production,  CP
  violation in the decay vertex being suppressed by the amount of the
 SM $t$-polarization. In order to study CP violation in the decay more
    useful appears  the  difference between the partial decay rates of
 $t$ and $\bar t$~\cite{{ECMF3},{Grz-decay}}: 
\be 
A_{CP} \equiv \frac{\Gamma \left(t
\rightarrow bW^+ \right) - \Gamma \left(\bar{t} \rightarrow \bar{b}W^-
             \right)}{\Gamma \left(t \rightarrow bW^+ \right) + \Gamma
 \left(\bar{t} \rightarrow \bar{b}W^- \right)} \, .\label{CPdecay} 
\ee
  This difference propagates into the differencies of the total number
       in events of the CP conjugate processes (\ref{b}) and these  of
                                                            (\ref{l}).

       In general the CP violating pieces in  the production and decay
  verticies have contributions from both  real and absorptive parts of
          the amplitude.  In accordance with this we have two types of
    observables: CP violation in the absorptive parts of the amplitude
($\Im m\, d^{\gamma ,Z}, \Im m\,f_L^{CP}, \Im m\,g_R^{CP}$) enter  the
     energy and the angular distributions.  Such an observable is also
   $A_{CP}$,  which  is proportional to the absorptive parts of the CP
 violating contributions of the $tbW$ vertex.  In order to measure  CP
      violation in the real parts of the amplitude one has to consider
    triple correlations of the type
\be  
{\cal T} = ({\bf q_1q_2q_3})
    \equiv ({\bf q_1 \times q_2 \cdot q_3})\, \label{triple} 
\ee 
where
   ${\bf q_{1,2,3}}$ can be any one of  the 3--momenta in each of  the
  processes (\ref{b}) or (\ref{l}). Triple product correlations 
 of particle momenta and spin for a
   general study of $CP$ violation in $t\bar t$ production in $e^+e^-$
                annihilation and in $pp$ collisions  have
 been proposed    in~\cite {{Bern}, {ECMF1,2}}.
 The  correlations (\ref{triple}) are
  called T-odd   as they change sign under the a flip of the 3-momenta
   involved.  However, this does not imply $T$, or through $CPT$ also 
$CP$ violation. The time reversal operation $T$ implies not only 
reverse of the particle momenta and spin, but also interchange of
 the initial and final states.  
 When loop corrections are included  the correlations ${\cal T}$ 
    can arise either from absorptive $CP$ invariant parts in the 
amplitude (so-called
 final state interactions~\cite{Henley}), or from $CP$ violation.  The
   former effect is a consequence of the unitarity of the $S$--matrix
 and it is our background. 
 It can be eliminated either by taking the difference between the
process we are interested in and its $CP$ conjugate or
  by direct estimates. $T$--odd correlations in the SM due to gluon or
Higgs boson exchange in the final states have  been considered
 in~\cite{{NP},{Sehgal-1}}.

 The contributions of the real and imaginary parts of the CP violating 
 form factors  to
           different pieces of the cross section  can be understood as
   follows. The CP violating terms  in the cross section come from the
        interference of the (real)  tree level SM amplitude and the CP
 violating loop corrections.  The i from the CP violating coupling and
    the i from the absorptive part of the loop  guarantee that the
        energy and angular distributions are real. The triple products
               (\ref{triple}) originate from the covariant quantity $i
  \varepsilon_{\alpha\beta\gamma\delta}p_1^\alpha p_2^\beta p_3^\gamma
p_4^\delta$ ($p_i$ is any of the 4-vectors of (\ref{b}) and (\ref{l}))
     when written in the laboratory frame. The i from the CP violating
                                        coupling and the i in front of
$\varepsilon_{\alpha\beta\gamma\delta}$ quarantee real contribution to
   the cross section only if real parts of the loop corrections
 are involved.

The presented formula for the distributions and the asymmetries are general 
and model independent. At the end we present numerical results in the
 Minimal Supersymmetric Standard Model (MSSM) with complex constants,
 where the CP violating form factors appear at one loop level.
 We have used the results of~\cite{dipole}, where  a complete analysis of 
 $d^\gamma$ and $d^Z$ in the MSSM was performed.

\section{The Formalism}

  In order to obtain analytic expressions for the distributions of the
decay products in the sequential processes  (\ref{b}) and (\ref{l}) we
         follow the formalism of \cite{BG}.  This approach allows 
 a clear physical interpretation of the different contributions in 
the cross section in terms of the polarization vectors of the
 decaying particle. For the cross sections of
        (\ref{b}) in the narrow width approximation  for the top quark
($\Gamma_t \ll m_t$)  we write 
 \be
  d\,\sigma^{b}_{\lambda\lambda'}
                                     = d\,\sigma^{t}_{\lambda\lambda'}
  \:\frac{d\,\Gamma_{\vec{t}}}{\Gamma_{t}} \,\frac{E_{t}}{m_{t} } \: ,
                        \qquad d\,\sigma^{\bar{b}}_{\lambda\lambda'} =
                                 d\,\sigma^{\bar{t}}_{\lambda\lambda'}
                        \:\frac{d\,\Gamma_{\vec{\bar{t}}}}{\Gamma_{t}}
                                       \,\frac{E_{\bar{t}}}{m_{t} }\,.
\label{generalb} 
\ee
 For the cross sections of (\ref{l}) in the narrow width approximation
   for $t$ and  $W$  ($\Gamma_t \ll m_t  , \Gamma_W \ll m_W$) we have
\be 
&&  d\sigma^{l^+}_{\lambda\lambda'}  = d\sigma^t_{\lambda\lambda'}
                        \, d\Gamma_{\vec {t}}\,\frac {E_t} {m_t \Gamma
              _t}\,d\Gamma_{\vec{W}^+}\frac {E_{W^+}}{m_W \Gamma_W}\,,
\label{l+} \nn 
                  &&  d\sigma^{l^-}_{\lambda\lambda'}  = d\sigma^{\bar
     t}_{\lambda\lambda'}  \, d\Gamma_{\vec {\bar t}}\, \frac {E_{\bar
          t}}{m_t \Gamma _t}\, d\Gamma_{\vec{W}^-} \frac {E_{W^-}}{m_W
            \Gamma_W}\,.\label{generall}  
\ee
 Here  $d\,\sigma^{t(\bar
      t)}_{\lambda\lambda'}$ is the differential cross section for $t$
  ($\bar t$) production in  $e^+e^-$ annihilation, $\lambda ,\lambda'$
      are the degrees of longitudinal polarization of the $e^-$, $e^+$
      beams.   $d\Gamma_{\vec{t}}$  ($d\Gamma_{\vec{\bar t}}$)  is the
           differential decay rate for $t \rightarrow b W^+$  ($\bar t
  \rightarrow \bar b W^-$) when the $t$ ($\bar t$) quark is polarized,
          its  polarization  determined in the previous 
production process, and
     $d\Gamma_{\vec {W}^\pm}$ is the differential decay rate of $W^\pm
      \rightarrow l^\pm\nu $, with the polarization states for $W^\pm$
determined in the preceeding $t$ ($\bar t$) decay,  $E_{t (\bar t)}$
 and
       $E_{W^\pm}$ are the energies of $t$ ($\bar t$) and $W^\pm$. All
        quantities are  in the c.m.system of $e^+ e^-$. $\Gamma_t$ and
           $\Gamma_W$ are the total decay widths of $t$ and $W$.  From
 (\ref{generalb}) and (\ref{generall})  the cross sections of
 (\ref{b}) and (\ref{l}) are obtained in terms of the polarization 
4-vector $\xi^t$
                        ($\xi^{\bar t}$) of the $t$ ($\bar t$) quarks.

         For the differential cross sections of (\ref{b}), assuming CP
  violation both in the production and decay vertices of the top-quark
  we obtain~\cite{angle}: 
\be  
 d\,\sigma^{b(\bar b)}_{\lambda\lambda'}
&=& \sigma^{b(\bar b)}_{0}  \left\{ A_{SM}^{b(\bar b)}+ A_d^{b(\bar b)} +
A_{g_R}^{b(\bar b)}\right\}\:d\cos\theta_{t(\bar t)}\:d\Omega_{b(\bar b)}
\label{sigmab}\\
    A_{SM}^{b(\bar b)} & =& 1\pm \alpha_b m_t\frac{(\xi_{SM}^{t(\bar t)}
   p_{b(\bar b)})}{(p_tp_b)}\, ,\label{ASMb}\\
 A_d^{b(\bar b)} & = &  \pm\alpha_b
       m_t \frac{(\xi_{CP}^{t(\bar t)} p_{b(\bar b)})}
{(p_tp_b)}\,, \label{ADb}\\
       A_{g_R}^{b(\bar b)} & = & \mp 2\left[ \Im m\, f_L^{CP} + \Im m\,
     g_R^{CP} \frac{m_t(m_t^2 -m_W^2)}{m_W(m_t^2 + 2m_W^2)}\right]-\nn
          &&-2\alpha_b m_t \left[  \Im m\, f_L^{CP} + \Im m\, g_R^{CP}
                 \frac{2m_t(m_t^2 -m_W^2)}{m_W(m_t^2 - 2m_W^2)}\right]
\frac{(\xi^{t(\bar t)}_{SM} p_{b(\bar b)})}{(p_tp_b)} \label{AgRb}
\ee 
Here   $ \sigma^{b(\bar
  b}_{0} $ is the cross section of  (\ref{b}) for unpolarized decaying
     top quarks,  $A_{SM}^{b(\bar b)}$ is the  contribution from the SM
       $t$-quark polarization,   $A_d^{b(\bar b)}$  describes the  $CP$
 violating pieces  due to $d^{\gamma ,Z}$, and $A_{g_R}^{b(\bar b)}$ --
 those   due to CP violation in the decay.  We consider the SM at tree
          level, which implies that  the SM form factors $f_L^{SM}$ and
      $g_R^{SM}$ are neglected.  We use a reference frame in which the
z-axis points the direction of $\bf q_e$, $\bf q_e$ and $\bf p_{t(\bar
        t )}$ determine the xz- plane, $\cos\theta_{t(\bar t)}$ is the
scattering angle of $t$ ($\bar t$).  The polarization vectors  $\xi^t$
         and $\xi^{\bar t}$  get contributions from the SM and from CP
    violating interactions: 
\be 
 \xi^{t(\bar t)} =\xi_{SM}^{t(\bar t)} +\xi_{CP}^{t(\bar t)}\,. 
\ee

For the differential  cross section of (\ref{l}) we obtain:
 \be 
d\sigma^{\pm}& =& \sigma_0^{{\pm}}
       \left\{ A_{SM}^{\pm}+ A_d^{\pm} + A_{g_R}^{\pm}\right\}\:
          d\cos\theta_{t(\bar t)}\:d\Omega_{b(\bar b)}\: d\Omega_{l^\pm}
\label{sigmal}\\
 A_{SM}^{\pm} & =& 1\mp\alpha_l m_t\frac{(\xi_{SM}^{t(\bar t)} p_{l^\pm})}
{(p_{t(\bar t)}p_{l^{\pm}})}\,
 ,\label{ASMl}\\ 
A_d^{\pm} & = & \mp\alpha_l m_t\,\frac{(\xi^{t(\bar
 t)}_{CP} p_{l^\pm})} {(p_{t(\bar t)}p_{l^{\pm}})}\,, \label{ADl}\\ 
 A_{g_R}^{\pm} & =&\mp 2\alpha_l\, \left[\Im m\,f_L^{CP} + 
\Im m\,g_R^{CP}\, \frac{m_t}{m_W}(1-\frac{m_W^2}
{2(p_{t(\bar t)} p_{l^\pm})})\right]\nn
&&-2\alpha_l\, \Im m\,g_R^{CP}\,
 \frac{(\xi_{SM}^{t(\bar t)}p_{b(\bar b)})}{m_W}\nn
&&+2\alpha_l m_t\, \left[\Im m\,f_L^{CP} + \frac{(p_tp_b)}{m_tm_W}\,
\Im m\,g_R^{CP}\right]\,
 \frac{(\xi_{SM}^{t(\bar t)}p_{l^{\pm}})}{(p_{t(\bar t)}p_{l^{\pm}})}\nn
    && \pm 2\alpha_l\,\Re e\, g_R^{CP}\, 
\frac{\varepsilon (\xi^{t(\bar t)}_{SM} p_{t(\bar t)}
 p_{l^\pm} p_{b(\bar b)})}
{m_W (p_{t(\bar t)}p_{l^\pm})}\,. \label{AgRl}
 \ee
 The indicies $\pm$ correspond to $l^+$ and $l^-$ production. 
$\sigma_0^{\pm}$ is the tree level SM cross section of (\ref{l}) for 
unpolarized decaying top quarks, $A_{SM}^{\pm}$ is the  contribution
                from the SM polarization,  the terms $A_d^{\pm}$ and
   $A_{g_R}^{\pm}$ contain the  $CP$--violating correlations
         due to $d^{\gamma ,Z}$  and due to CP violation in the decay,
respectively.  
 The quantity $\varepsilon (p_1p_2p_3p_3)$ is abbreviation of
$\varepsilon_{\alpha\beta\gamma\delta} p_1^\alpha p_2^\beta p_3^\gamma
                                                          p_4^\delta$.

 Eqs. (\ref{sigmab}) and (\ref{sigmal}) are our basic formula. 
They imply that CP violation enters
       the angular and energy  distributions of the $b$ quarks and the
        leptons only  through the polarization of the top quarks.  The
   coefficients $\alpha_b$ and $\alpha_l$ determine the sensitivity of
 the $b$-quarks and the leptons  to the $t$-polarization.
 We have~\cite{Bern}:
 \be
       \alpha_{b}  = \frac{m_{t}^2 - 2 m_{\scriptscriptstyle W}^{2}}
{m_{t}^{2} + 2 m_{\scriptscriptstyle W}^{2}}\,,\qquad  \alpha_l = 1\:.
      \ee
 The sensitivity to CP violation  in the production plane  is
   determined  by $\alpha_b$ or $\alpha_l$ and $\xi_{CP}$.
 The sensitivity to CP
          violation in the decay is determined by $\alpha_b$ or $\alpha_l$, 
 the  form factors $f_L^{CP}$, $g_R^{CP}$, and the SM top quark polarization
   $\xi_{SM}$.  Thus,  the sensitivity to $f_L^{CP}$ and $g_R^{CP}$ 
in the distribution of the decay products is
 suppressed  by the amount of the SM $t$-polarization. 
 In the next section we shall
                  give the  explicit expressions for $\xi^{t,\bar t}$.
 The first terms in $A_{g_R}^{b \bar b)}$ and $A_{g_R}^{\pm}$ are 
independent on the top-polarization, which implies  
that CP violation in the decay will enter the total cross sections. 

  $\sigma^{b}_{0}$ and $\sigma^{l}_{0}$  determine the differential SM
    cross  sections of (\ref{b}) and (\ref{l}) for totally unpolarized
          decaying top quarks with  longitudinally polarized initial
electron--positron beams. We have: 
\be 
 &&  \sigma^{b(\bar{b})}_0  =
           \alpha_{em}^2 \:\frac{3 \beta}{2 s}  \:
\frac{\Gamma_{t\to bW}}{\Gamma_t}  \:
\frac{m_t^2 E_{b(\bar{b})}^2}{( m_t^2 -m_{\scriptscriptstyle W}^2 )^2}
                                    \:N^{t(\bar{t})}_{\lambda\lambda'}
\label{sigmab0} \\
                         &&\sigma_0^{\pm} =
   \frac{\alpha^4_{em}}{\sin^4\Theta_W}\frac{3\beta}{32 \pi s}
 \frac{\left[m_t^2-2(p_{t(\bar t)}p_{l^\pm})\right]
(p_{(t\bar t)}p_{l^\pm})}
{ m_t\Gamma_t\,m_W\Gamma_W}
\frac{E_{b(\bar b)}^2}{m_t^2-m_W^2}\frac{E_{l^\pm}^2}{m_W^2}
N^{t(\bar   t)}_{\lambda\lambda'} \,. \label{sigmal0} 
\ee 
 Here $E_{b(\bar b)}$
   and $E_{l^\pm}$ are the energies of  the b-quarks and the final
                    leptons in the c.m.system:
 \be
        E_b=\frac{m_t^2-m_W^2}{2E}\frac{1}{1-\beta \cos\theta_{tb}}\,
,\qquad   E_{\bar
    b}=\frac{m_t^2-m_W^2}{2E}\frac{1}{1-\beta  \cos\theta_{\bar t\bar b}}
  \ee
 \be 
 E_{l^+}=\frac{m_W^2}{2\left[
             E(1-\beta \cos\theta_{tl^+})
-E_b(1-\cos\theta_{bl^+})\right]}\,, 
 \ee
\be 
 E_{l^-}=\frac{m_W^2}{2\left[
             E(1-\beta \cos\theta_{\bar tl^-})
-E_{\bar b}(1-\cos\theta_{\bar bl^-})\right]}\,, 
 \ee
 $\sqrt{s}$ is the total c.m.energy,
   $\beta = \sqrt{1 - 4 m_{t}^{2} / s}$ is the  velocity of the $t$
     quark, 
 \be 
 \cos\theta_{tb}  = \frac{({\bf p}_t \!\cdot\!
    {\bf p}_{b})} {|{\bf p}_t|\,|{\bf p}_b|}  = \sin\theta_{t}
   \sin\theta_{b} \cos\phi_{b}  + \cos\theta_{t} \cos\theta_{b}\,, 
\quad {\rm etc}.
   \ee 
 We take $m_{b} = 0$,.  $\Gamma_{t\to bW}$ is the partial decay
   width of the top quark for the decay $t\to bW$:
 \be 
 \Gamma (t\to
                 bW)= \frac{G_F m_t^3}{8\sqrt 2 \pi}
   \left(\frac{m_t^2-m_W^2}{m_t^2}\right)^2 \frac{m_t^2 + 2m_W^2} {
        m_t^2}\vert V_{tb}\vert^2\,, 
\ee 
where $V_{tb}$ is the
   corresponding element in the CKM mixing matrix.  In (\ref{sigmab0})
   and (\ref{sigmal0}) we take $\Gamma_{t\rightarrow bW}/\Gamma_t =
   1$.  We use the notation: 
\be 
N^{t(\bar{t})}_{\lambda\lambda'}  = (
      1 + \beta^{2} \cos^{2} \theta_{t(\bar{t})} ) F_{1} + ( 1 -
   \beta^{2} ) F_{2}  \plmin 2 \beta \cos \theta_{t(\bar{t})} \, F_{3}
   \; .  
\ee 
 The dependence on the beam polarizations comes through
   the  functions $F_{i}$, $i=1,2,3$, given by  
\be 
 F_{i}  = ( 1 -
   \lambda \lambda' ) F_{i}^{0}  + ( \lambda - \lambda' ) G_{i}^{0}
\label{F} 
         \ee  
where 
 \be
 \hspace{-5mm} F_{1,2}^{0}  &=& \sfrac{4}{9} -
         \sfrac{4}{3} c_{V} g_{V} h_{Z}  + ( c_{V}^{2} + c_{A}^{2} ) (
    g_{V}^{2} \pm g_{A}^{2} ) h_{Z}^{2} \nn 
G_{1,2}^{0} &=& 
       \sfrac{4}{3} c_{A} g_{V} h_{Z}  + 2 c_{V} c_{A} ( g_{V}^{2} \pm
g_{A}^{2} ) h_{Z}^{2} \nn
F_{3}^{0}  &=& 
g_{A} h_{Z}  + 4 c_{V} c_{A} g_{V} g_{A} h_{Z}^{2} \nn 
G_{3}^{0} &=& 
 - \sfrac{4}{3} c_{V} g_{A} h_{Z}  + 2 ( c_{V}^{2} + c_{A}^{2} )
                g_{V} g_{A} h_{Z}^{2}
\label{FG0}
  \ee 
The quantities  $c_{V} = - (1/2) + 2
 \sin^{2}\Theta_{\scriptscriptstyle W}$,  and $c_{A} = (1/2)$  are the
SM couplings of $Z$ to the electron,  $h_{Z} = [ s / ( s - m_{Z}^{2} )
                       ]  / \sin^{2} 2 \Theta_{\scriptscriptstyle W}.$

\section{The polarization vector}

  The amplitude for $e^{+}e^{-} \to t\bar{t}$, assuming  CP violation,
       is 
\be 
{\mathcal M}= i \frac{e^{2}}{s}  \bar{v}{ (q_{\bar{e}})}
           \gamma_{\mu}  u{ (q_{e})} ({\mathcal V}^{\gamma})^{\mu} - i
 \frac{g_{Z}^{2}}{s - m_{Z}^{2}} \bar{v}{ (q_{\bar{e}})}  \gamma_{\mu}
 ( c_{V} + c_{A} \gamma^{5} )  u{ (q_{e})} ({\mathcal V}^{Z})^{\mu}\,.
\label{amplitude}
 \ee

  Now  we will give the expressions \cite{angle}
 for the polarization four--vectors
  $\xi_{\mu}^{t}$ and  $\xi_{\mu}^{\bar t}$ 
  including the dependence
on the electric and weak dipole moment form factors.  $\xi^{t,\bar t}$
            determine the spin density matricies of the decaying 
$t$ and $\bar t$ quarks: 
\be 
 \rho (p_t ) =
     \frac{1}{2}(1+{\rlap/\xi}^t \gamma^5)\Lambda (p_t)\,, \qquad \rho
 (-p_{\bar t} ) =- \frac{1}{2}(1+{\rlap/\xi}^{\bar t} \gamma^5)\Lambda
             (-p_{\bar t})\,,\nonumber 
\ee
\be
 \Lambda(p_t)=\Sigma_r u_r(p_t){\bar
                                     u}_r(p_t)=({\rlap/p}_t + m_t)\,.
\nonumber
 \ee

   As $(p_{t}\xi)=0$, in general the polarization vector $\xi_{\mu}^t$
can be decomposed covariantly along three independent 
four--vectors orthogonal  to
   $p_{t}$: two of them, $Q_{e}^{\mu}$ and $Q_{\bar{e}}^{\mu}$  are in
the production plane: 
 \be 
 Q_{e}^{\mu}  = q_{e}^{\mu} - \frac{( p_{t}
      q_{e} )}{m_{t}^{2}} p_{t}^{\mu} \; , \qquad  Q_{\bar{e}}^{\mu} =
          q_{\bar{e}}^{\mu}  - \frac{( p_{t} q_{\bar{e}} )}{m_{t}^{2}}
                p_{t}^{\mu} \;  
\ee 
and the third one is normal to it:
     $\varepsilon_{\mu\alpha\beta\gamma}  p_{t}^{\alpha} q_{e}^{\beta}
            q_{\bar{e}}^{\gamma}$.  Most generally, we can write:  
\be
         \xi_{\mu}^t  = P_{e}^{t} ( Q_{e} )_{\mu}  + P_{\bar{e}}^{t} (
       Q_{\bar{e}} )_{\mu}  + D^{t} \varepsilon_{\mu\alpha\beta\gamma}
                p_{t}^{\alpha} q_{e}^{\beta} q_{\bar{e}}^{\gamma} \: .
\label{xit}
   \ee 
The components $P_{e(\bar{e})}^{t}$ get contributions from both
           SM and {CP violating} terms. The SM at tree level  does not
          contribute to the normal component $D^t$. Thus we have:  
\be
                    P_{e(\bar{e})}^{t} = P_{e(\bar{e})}^t({\it SM})  +
             P_{e(\bar{e})}^t({\scriptstyle CP} )  \: , \qquad D^{t} =
                                            D^t({\textstyle CP} ) \: .
\label{SMCP} 
               \ee  
The polarization 4-vector is determined by the
        expression~\cite{BG} 
 \be  \xi_{\mu}^t  = \left( g_{\mu\nu} -
       \frac{p_{t\mu} p_{t\nu}}{m_t^2}\right) \frac{ \Tr[ {\mathcal M}
        \bar{\Lambda} (p_{\bar{t}}) \bar{{\mathcal M}} \Lambda (p_{t})
           \gamma^{\nu} \gamma^{5} ]} {\Tr[ {\mathcal M} \bar{\Lambda}
         (p_{\bar{t}}) \bar{{\mathcal M}} \Lambda (p_{t}) ]}
 \ee 
where
  ${\mathcal M}$ is the amplitude (\ref{amplitude}). The projection
   operator $( g_{\mu\nu} - m_{t}^{-2} p_{t\mu} p_{t\nu} )$ guarantees
the condition ($\xi p_t)=0$.  In the c.m.system the SM contribution to
$P_{e(\bar{e})}^t({\it SM})$ at  tree-level is 
\be 
 P_{e}^t({\it SM})
     &=& \frac{2 m_{t}}{s} \frac{1}{N^{t}_{\lambda\lambda'}}   [ ( 1 -
 \beta \cos\theta_{t} ) ( G_{1} - G_{3} ) + ( 1 + \beta \cos\theta_{t}
                                                             ) G_{2} ]
\label{PSMe}
                   \\  P_{\bar{e}}^t({\it SM}) &=& - \frac{2 m_{t}}{s}
   \frac{1}{N^{t}_{\lambda\lambda'}}  [ ( 1 + \beta \cos\theta_{t} ) (
                G_{1} + G_{3} ) + ( 1 - \beta \cos\theta_{t} ) G_{2} ]
\label{PSMebar}
       \ee
 where $G_{i}$, $i=1,2,3$ are given by:  
\be  
G_{i}  = ( 1 -
      \lambda \lambda' ) G_{i}^{0}  + ( \lambda - \lambda' ) F_{i}^{0}
\label{G} 
 \ee 
 with $F_{i}^{0}$ and $G_{i}^{0}$ as defined in (\ref{FG0}).  The
  CP violating dipole moment form factors $\dgs$ and $\dzs$ induce two
   types of contributions: due to their real and  imaginary parts. The
absorptive parts $\mIm \dgz$ contribute to  $P_{e(\bar{e})}^t({\it CP})$:  
\be 
 P_{e}^t({\it CP} ) &=& - \frac{2}{m_{t}}
      \frac{1}{N^{t}_{\lambda\lambda'}} [ ( 1 + \beta \cos\theta_{t} -
  \beta^{2} \sin^{2}\theta_{t} )  \mIm H_{1}  \nn & & \hspace{2cm} - (
                       \beta \cos\theta_{t} + \beta^{2} ) \mIm H_{2} ]
\label{PCPe} 
                     \\ P_{\bar{e}}^t({\it CP} ) &=& - \frac{2}{m_{t}}
     \frac{1}{N^{t}_{\lambda\lambda'}}  [ ( 1 - \beta \cos\theta_{t} -
  \beta^{2} \sin^{2}\theta_{t} )  \mIm H_{1}  \nn & & \hspace{2cm} - (
                  \beta \cos\theta_{t} - \beta^{2} ) \mIm H_{2} ] \; .
\label{PCPebar}
      \ee 
Here we have used the notation: 
 \be 
 H_{i}  = ( 1 - \lambda
              \lambda' ) H_{i}^{0}  + ( \lambda - \lambda' ) D_{i}^{0}
\label{H} 
  \ee 
 where  
\be 
 H_{1}^{0}  &=& ( \sfrac{2}{3} - c_{V} g_{V} h_{Z} )
  \dgs - ( \sfrac{2}{3} c_{V} h_{Z}  - ( c_{V}^{2} + c_{A}^{2} ) g_{V}
   h_{Z}^{2} ) \dzs \nn
 H_{2}^{0}  &=& 
    h_{Z} \; \dgs  + 2 c_{V} c_{A} g_{A} h_{Z}^{2} \;
\dzs \nn 
D_{1}^{0} &=&  - c_{A} g_{V} h_{Z} \; \dgs -
  ( \sfrac{2}{3} c_{A} h_{Z}   - 2 c_{V} c_{A} g_{V}
    h_{Z}^{2} ) \dzs \nn
 D_{2}^{0} &=&  - c_{V} g_{A}
     h_{Z} \; \dgs   + ( c_{V}^{2} + c_{A}^{2} ) g_{A}
                                                     h_{Z}^{2} \; \dzs
\label{H0}
         \; .  \ee
 The real parts of $\dgz$ determine the CP violating
         contribution  $D^t({\it CP} )$ to the normal component of the
          polarization vector:  
\be
   D^t({\it CP} ) &=&
    \frac{8}{m_{t} s} \frac{1}{N^{t}_{\lambda\lambda'}} [ \eRe D_{1} +
                                     \beta \cos\theta_{t} \eRe D_{2} ]
\label{DCP} 
       \ee 
Here 
 \be  D_{i}  = ( 1 - \lambda \lambda' ) D_{i}^{0}  + (
                                        \lambda - \lambda' ) H_{i}^{0}
\label{D} 
  \ee 
 Note that $H_{i}^{0}$ are C--odd and P--even, while $D_{i}^{0}$
    are  C--even and P--odd functions of the coupling constants in the
       production process $e^{+}e^{-} \to t\bar{t}$. This implies that
      $H_{i}$ are C--odd and CP--odd, while $D_{i}^{0}$ are P--odd and
                                                   CP--odd quantities.

     The polarization four--vector $\xi^{\bar t}$ for the anti--top is
          obtained through C--conjugation. This leads to the following
  replacements in the expressions for $\xi_{\mu}^t$, $F_{i}$, $G_{i}$,
 $H_{i}$, and $D_{i}$: 
 \be
  p_{t} \to p_{\bar{t}} \, , \: (2/3) e \to
-  (2/3) e \, , \: g_{V} \to - g_{V} \, , \: \dgz \to - \dgz \: .  
\ee
    We have:  
\be 
 \xi_{\mu}^{\bar t}  = P_{e}^{\bar{t}} ( \bar{Q}_{e}
        )_{\mu} + P_{\bar{e}}^{\bar{t}} ( \bar{Q}_{\bar{e}} )_{\mu}  +
  D^{\bar{t}} \varepsilon_{\mu\alpha\beta\gamma}  p_{\bar{t}}^{\alpha}
                               q_{e}^{\beta} q_{\bar{e}}^{\gamma} \: .
\label{xitbar}
           \ee 
where  \be  \bar{Q}_{e}^{\mu}  = q_{e}^{\mu}  - \frac{(
         p_{\bar{t}} q_{e} )}{m_{t}^{2}} p_{\bar{t}}^{\mu} \; , \qquad
    \bar{Q}_{\bar{e}}^{\mu} = q_{\bar{e}}^{\mu}  - \frac{( p_{\bar{t}}
   q_{\bar{e}} )}{m_{t}^{2}} p_{\bar{t}}^{\mu} \; . 
 \ee 
In analogy to
              eq.(\ref{SMCP}) we define: 
 \be 
 P_{e(\bar{e})}^{\bar{t}}
 =P_{e(\bar{e})}^{\bar{t}}({\it SM})  + P_{e(\bar{e})}^{\bar{t}} ({\it
            CP} ) \: , \qquad D^{\bar{t}} =D^{\bar{t}}({\it CP} ) \: .
\label{barSMCP}
   \ee 
and obtain:  
\be
  P_{e}^{\bar{t}}({\it SM})  &=&
   \frac{2 m_{t}}{s} \frac{1}{N^{t}_{\lambda\lambda'}}   [ ( 1 - \beta
                \cos\theta_{\bar{t}} ) ( G_{1} + G_{3} ) + ( 1 + \beta
\cos\theta_{\bar{t}} ) G_{2} ] \\ 
P_{\bar{e}}^{\bar{t}}({\it SM})  &=&
-  \frac{2 m_{t}}{s} \frac{1}{N^{t}_{\lambda\lambda'}}   [ ( 1 + \beta
                \cos\theta_{\bar{t}} ) ( G_{1} - G_{3} ) + ( 1 - \beta
  \cos\theta_{\bar{t}} ) G_{2} ] \\ 
 P_{e}^{\bar{t}} ({\it CP} ) &=& -
       \frac{2}{m_{t}} \frac{1}{N^{t}_{\lambda\lambda'}} [ ( 1 + \beta
     \cos\theta_{\bar{t}}  - \beta^{2} \sin^{2}\theta_{\bar{t}} ) \mIm
H_{1}  \nn 
& & \hspace{2cm} + ( \beta \cos\theta_{\bar{t}} + \beta^{2}
              ) \mIm H_{2} ] \\ P_{\bar{e}}^{\bar{t}} ({\it CP} )&=& -
       \frac{2}{m_{t}} \frac{1}{N^{t}_{\lambda\lambda'}} [ ( 1 - \beta
\cos\theta_{\bar{t}} - \beta^{2} \sin^{2}\theta_{\bar{t}} ) \mIm H_{1}
\nn & & \hspace{2cm} + ( \beta \cos\theta_{\bar{t}} - \beta^{2} ) \mIm
               H_{2} ] \\ D^{\bar{t}}({\it CP} ) &=& \frac{8}{m_{t} s}
                \frac{1}{N^{t}_{\lambda\lambda'}} [ \eRe D_{1} - \beta
                                     \cos\theta_{\bar{t}} \eRe D_{2} ]
\label{DbarCP} 
\ee

 The expressions for the cross sections of (\ref{b}) and (\ref{l}) are
   naturally expressed in terms of the dimensionless combinations: 
\be
&&  {\cal P}^{t(\bar t)}_{\pm} =\frac{s}{m_t}( P^{t(\bar t)}_{e} \pm
  P^{t(\bar t)}_{\bar{e}}) = {\cal P}_\pm^{t(\bar t)}({\it SM}) +  {\cal
 P}_\pm^{t(\bar t)}({\it CP} )\,,\nn
&& {\cal D}^{t(\bar t)}=sm_t  D^{t(\bar t)}= 
 {\cal D}^{t(\bar t)}({\it CP})\label{calPD}
\ee

\section{The process $e^+ e^- \rightarrow bX$}
\subsection{The differential cross section}  

In this section we summarize the rezults of 
~\cite{angle} and ~\cite{energy}. 
 Using the explicit expressions eqs.(\ref{xit}) and (\ref{xitbar})  for
 the top and the anti--top quark polarization four--vectors we  obtain
   from (\ref{sigmab}) the analytic formula for the  cross sections of
                        (\ref{b}) in the c.m.system~\cite{angle}:  
\be
   d\,\sigma^{b(\bar{b})}_{\lambda\lambda'} &=&  \sigma^{b(\bar{b})}_0
     {\scriptstyle (\lambda,\lambda')} \left\{   1 \plmin   \alpha_{b}
            \frac{m_{t}^2} {m_{t}^{2} - m_{\scriptscriptstyle  W}^{2}}
       \frac{E_{b(\bar{b})}}{\sqrt s} \left[ {\cal P}^{t(\bar{t})}_{+}
 \left( 1 - \frac{1 - \beta \cos \theta_{tb(\bar{tb})}}{1 - \beta^{2}}
          \right) \right.\right. \hspace{-15mm} \nn
  & & \hspace{15mm}
                     \left.\left.  - {\cal P}^{t(\bar{t})}_{-}  \left(
   \cos\theta_{b(\bar{b})}  - \beta \cos \theta_{t(\bar{t})} \frac{1 -
              \beta \cos \theta_{tb(\bar{tb})}}{1 - \beta^{2}} \right)
  \right.\right. \hspace{-15mm} \nn 
 & & \hspace{15mm} \left.\left.  +
             \frac{s \beta}{2m_t^2} {\cal D}^{t(\bar{t})}
                     (\hat{\mathbf{q}}_{e} \hat{\mathbf{p}}_{t(\bar{t})}
        \hat{\mathbf{p}}_{b(\bar{b})}) \right] \right\} \: d\,
 \cos\theta_{t(\bar{t})} \: d\,\Omega_{b(\bar{b})} \; ,\label{dsigmab}
       \ee 
where 
 $\hat{\mathbf{q}}$ and  $\hat{\mathbf{p}}$ are  unit
3--vectors  in the direction of the particles. $\sigma^{b(\bar{b})}_0$
 is given in eq.(\ref{sigmab0}).  In (\ref{dsigmab}) we have kept only
       the dependence on the electroweak dipole moment form factors and
                 neglected CP violation in the decay of the top quark.


\subsection{The angular distributions of $b$ and $\bar b$ quarks}
                    Integrating (\ref{dsigmab})  over 
$\cos\theta_{t(\bar t)}$  and $\varphi_{b(\bar b)}$ 
  we
obtain the $\cos\theta_{b(\bar b)}$--distribution  of
          the $b$($\bar{b}$) quarks in the c.m.system:  \be  \frac{d\,
 \sigma^{b(\bar{b})}_{\lambda\lambda'}} {d\, \cos \theta_{b(\bar{b})}}
    &=& \frac{3 \pi \alpha_{em}^{2} \beta}{2 s}  \frac{\Gamma_{t \to b
  W}}{\Gamma_{t}}  \left( a_{0}^{b(\bar{b})} \plmin a_{1}^{b(\bar{b})}
                \cos \theta_{b(\bar{b})} + a_{2}^{b(\bar{b})} \cos^{2}
                                           \theta_{b(\bar{b})} \right)
\label{angle} 
     \ee  
where  
\be a_i^b= a_i^{SM} + a_i^{CP}\,,\qquad a_i^{\bar b}=
                                a_i^{SM} - a_i^{CP}\nonumber  
\ee 
\be a_{0}^{SM}
= ( 1 + \beta^{2} - b ) F_{1} + ( 1 - \beta^{2} ) F_{2} - \alpha_{b} (
   b - \beta^{2} ) G_{3},\,\, a_{0}^{CP} = - 2 \alpha_{b} b \mIm H_{1}
\nonumber                                                    \ee
 \be 
a_{1}^{SM}
 = 2 b F_{3} - \alpha_{b} \left(  ( 1 + \beta^{2} - 2 b ) G_{1}  + ( 1
  -          \beta^{2} ) G_{2}  \right) ,\,\, \,\,    a_{1}^{CP} = - 4
                           \alpha_{b} b \mIm H_{2}\nonumber  
\ee
 \be 
a_{2}^{SM}
       =  ( 3 b - 2 \beta^{2} ) F_{1} + 3 \alpha_{b} ( b - \beta^{2} )
     G_{3},\quad\qquad\qquad  a_{2}^{CP} =   6 \alpha_{b} b \mIm H_{1}
                                           \nonumber        \ee 
\be
 b
      = 1  - \sfrac{1 - \beta^{2}}{2 \beta} \ln[ \sfrac{1 + \beta}{1 -
    \beta} ]  \; .  \nonumber
\ee 
The two independent combinations of the dipole
        moment form factors  $H_1$ and $H_2$ enter (\ref{angle}), which
   implies that studying the angular distribution of  the $b$ and $\bar
     b$ quarks one can obtain information about  both  $\Im md^\gamma$
                                                       and $\Im md^Z$.

   These formulae coincide with the analogous SM expressions  obtained
  in~\cite{Sehgal-1} for the unpolarized $e^{+}e^{-}$  and 
with~\cite{Draganov} for polarized $e^{+}e^{-}$.

\subsection{The energy distributions of $b$ and $\bar b$ quarks}

     Using (\ref{dsigmab}) it is starightforward to obtain  the energy
     distribution  of the $b$ and $\bar{b}$ quarks if one moves to the
       frame where the  $z$--axis points into the direction of the top
                                 quarks~\cite{energy}: 
 \be 
 \frac{d\,
       \sigma^{b(\bar{b})}_{\lambda\lambda'}} {d\, x_{b(\bar{b})}} &=&
           \frac{\pi \alpha_{em}^{2}}{s}  \frac{m_{t}^{2}}{m_{t}^{2} -
            m_{\scriptscriptstyle W}^{2}}  \left( c_{0}^{b(\bar{b})} +
  c_{1}^{b(\bar{b})} x_{b(\bar{b})} \right) \label{energy} 
 \ee 
 where
 \be
 && c_i^b = c_i^{SM}+ c_i^{CP}\,,\qquad \qquad \quad  c_i^{\bar b}
       = c_i^{SM} - c_i^{CP}\nn
 &&  c_{0}^{SM}=  N_{tot} + 4 \alpha_{b}
         G_{3}\,, \qquad\quad c_0^{CP} = 8\alpha_b \mIm H_{1}\,, \nn
 &&
   c_{1}^{SM}= -8\alpha_b\frac{m_t^2}{m_t^2 -m_W^2}\,, \qquad c_1^{CP}
  =-16\alpha_b \frac{m_t^2}{m_t^2 -m_W^2} \mIm H_{1} \,.  
\nonumber
\ee 
 We have
 used the conventional dimensionless energy variables  $x_{b(\bar{b})}
    = \frac{2 E_{b(\bar{b})}}{\sqrt{s}}$ and the notation
 \be 
N_{tot}=
       (3+\beta^2) F_1 + 3(1-\beta^2) F_2\,.\label{Ntot}
  \ee 
Note that the linear
                behaviour of the spectra is introduced only by the top
          polarization -- both $c_1^{SM}$ and $c_1^{CP}$ are
  proportional to $\alpha_b$. This may serve as a 
good analyser of the spin of the top quark in SM~\cite{Draganov}.  
Studying the energy distributions one cannot obtain information about
 $d^\gamma$ and $d^Z$ independently -- only one combination $H_1$
enters the CP   violating terms $c^{CP}_{0,1}$ in
 (\ref{energy}).

\subsection{CP violating  asymmetries}

  The electroweak dipole moment form factors $\dgz$ have both real and
  imaginary parts. To obtain information about the dipole moment
 form factors from the differential cross section is a difficult 
task and it acquires very high precision of measurements.
 In the following we consider different integral observables, 
               sensitive to $\eRe \dgz$ and to $\mIm \dgz$ seperately.

\vspace{0.5cm}
\hspace{1cm}{\bf i)}
 $\mIm \dgz$ can be measured both by the angular-$\cos\theta_b$ 
and the energy asymmetries.

CP invariance for the angular distribution (\ref{angle})  implies: 
\be
      \frac{d\,\sigma^{b}_{\lambda\lambda'}(\cos\theta_b=\cos\theta )}
       {d\,\cos \theta_{b}} \:  = \frac{d\,\sigma^{\bar{b}}_{-\lambda'
-             \lambda} (\cos\theta_{\bar b}=\pi -\cos\theta)} {d\,\cos
 \theta_{\bar{b}}} \,. \label{CP-diff}  
\ee
 Note that in this equation
    and in the following ones, the first  lower index of $d\sigma^b$
            and $d\sigma^{\bar b}$  denotes the degree of longitudinal
     polarization of the  electron beam and the second one that of the
                                                        positron beam.

  Let $\sigma_F^{b (\bar b)}(\lambda ,\lambda' )$ and 
 $\sigma_F^{b (\bar b)}(\lambda ,\lambda' )$ denote the cross section 
of $b$ and $\bar b$
   produced in the forward and backward hemispheres, respectively. Let
              $A_{FB}^{b(\bar b)}(\lambda ,\lambda')$ is  the standard
       forward--backward asymmetries for the  $b$ and $\bar{b}$ quarks:  
\be
        A^{b(\bar{b})}_{FB} {\scriptstyle (\lambda,\lambda')} = \frac{
           \sigma^{b(\bar{b})}_{F} {\scriptstyle (\lambda,\lambda')} -
          \sigma^{b(\bar{b})}_{B} {\scriptstyle (\lambda,\lambda')}} {
           \sigma^{b(\bar{b})}_{F} {\scriptstyle (\lambda,\lambda')} +
      \sigma^{b(\bar{b})}_{B} {\scriptstyle (\lambda,\lambda')}}  \; ,
\label{AFB0}
             \ee
 Then we define the CP violating asymmetry  ${\mathcal
   A}^{FB}_{\lambda\lambda'}$: 
\be 
{\mathcal A}^{FB}_{\lambda\lambda'}
      =A_{FB}^{b}{\scriptstyle (\lambda,\lambda')} +{A}_{FB}^{\bar{b}}
       {\scriptstyle (-\lambda',-\lambda)} = -12\alpha_b b \frac{\Im m
                                       H_2}{N_{tot}}\;.\label{AFB} \ee
Other CP violating angular asymmetries, including also 
the general analytic expressions for their dependence on the experimental 
cuts are presented in \cite{angle}. 

   CP invariance for  the energy spectra of $b$ and $\bar b$
implies: 
\be 
\frac{d\sigma^{b}_{\lambda ,\lambda'}(x_b=x)}{d\,x_{b}} =
          \frac{d\sigma^{\bar{b}}_{-\lambda' ,-\lambda}(x_{\bar b}=x)}
  {d\,x_{\bar{b}}}\; .\label{E-diff} 
\ee
The corresponding integrated energy observable 
${\cal A}^E_{\lambda\lambda'}$ indicating 
CP violation is
\be 
{\cal A}^E_{\lambda\lambda'}
= R_{\lambda \lambda'}^b - R_{-\lambda' -\lambda}^{\bar b} =
= \frac{- 4 \alpha_{b} \beta \Im m H_{1}}{ N_{tot}} 
\; ,
\label{AE}
\ee
where
\be 
R_{\lambda\lambda'}^{b (\bar{b})}
=\frac{ N^{b(\bar{b})}(x > x_0,\lambda,\lambda') 
- N^{b(\bar{b})}(x < x_0,\lambda,\lambda')}
{N_{tot}^{b(\bar b )}(\lambda ,\lambda')}
\equiv \frac{\Delta N^{b(\bar{b})}(\lambda,\lambda')} 
 { N_{tot}^{b (\bar{b})}(\lambda,\lambda')}  
\; ,\nonumber
\ee
\be
x_{0} = \frac{x_{min} + x_{max}}{2},\quad
 x_{min} =\frac{2(m_t^2 - m_W^2)}{s(1+\beta )}, \quad 
x_{max} =\frac{2(m_t^2 - m_W^2)}{s(1-\beta )}\,,
\ee
$N^{b(\bar{b})}(x>x_0,\lambda,\lambda')$ is the number of $b$($\bar{b}$) 
quarks with  $x>x_0$ for  beam polarizations $\lambda$,  
 $\lambda'$, $N_{tot}^{b(\bar b)}(\lambda ,\lambda')$ is the total number
 of $b$ ($\bar b$) quarks (the total cross section) of (\ref{b}):
\be 
N_{tot}^b = N_{tot}^{\bar b} = \frac{\pi \alpha_{em}^2}{s} 
\,\beta\,\frac{\Gamma_{t\to bW}}{\Gamma_t} N_{tot},
\ee
 $N_{tot}$ is given by (\ref{Ntot}).

            The electroweak dipole moment form factors $\dgz$ enter two
      combinations $H_1$ and $H_2$. The asymmetries ${\cal A}^{FB}$ 
and ${\cal A}^E$
    provide two independent measurements of their imaginary parts and
 thus of $\Im md^\gamma$ and $\Im md^Z$. Through $H_1$ and $H_2$ these
        asymmetries  depend  on the beam polarization that can 
strongly
  enhance (or decrease)  the effects of CP violation we are interested
    in. Measurements performed with opposite beam polarizations can be
 used to disentangle $H_i^0$ from $D_i^0$.  In analogy to the standard
      forward-backward asymmetries (\ref{AFB0}) we define the following
     polarization asymmetries:   
\be 
P_{FB}^{b(\bar{b})} =
         \frac{( 1 - \lambda \lambda' )}{( \lambda - \lambda') } \cdot
              \frac{(\sigma^{b(\bar{b})}_{F} -\sigma^{b(\bar{b})}_{B})
                          (\lambda,\lambda') -(\sigma^{b(\bar{b})}_{F}
 -                      \sigma^{b(\bar{b})}_{B}) (-\lambda,-\lambda')}
                   {(\sigma^{b(\bar{b})}_{F} +\sigma^{b(\bar{b})}_{B})
                          (\lambda,\lambda') +(\sigma^{b(\bar{b})}_{F}
  +\sigma^{b(\bar{b})}_{B}) (-\lambda,-\lambda')}\; .  
\ee 
Then the CP
        violating asymmetry is 
\be 
 {\cal P}^{FB}  = P_{FB}^{b}  +
                     P_{FB}^{\bar{b}}  = - 12 \alpha_{b} b  \frac{\mIm
                                                 D_{2}^{0}}{N_{tot}^0}
\label{PFB}
           \ee
 where 
\be 
N_{tot}^{0} =  N_{tot} (\lambda=\lambda'=0) =
                            (3+\beta^2)F_1^0 + 3(1-\beta^2)F_2^0\,. 
\ee

 Analogously we define the polarization CP violating asymmetry for the
energy spectra: 
\be
{\cal P}^E 
= R_{P}^{b} - R_{P}^{\bar{b}}
= \frac{- 4 \alpha_{b} \beta \mIm D_{1}^{0}}{ N_{tot}^{0}} 
\; ,
\label{PE}
\ee
where
\be
R_{P}^{b(\bar{b})}
= \frac{( 1 - \lambda \lambda' )}{( \lambda - \lambda' )} \cdot 
\frac{\Delta N^{b(\bar{b})}(\lambda,\lambda')
     -\Delta N^{b(\bar{b})}(-\lambda,-\lambda')}
     {N_{tot}^{b(\bar{b})}(\lambda,\lambda')
     +N_{tot}^{b(\bar{b})}(-\lambda,-\lambda')} 
\; .
\ee

\vspace{0.5cm}
\hspace{1cm}{\bf ii)}
 The real parts of $\dgz$ can be
                              singled out by  measuring triple product
   correlations~\cite{{Bern},{ECMF1,2}}.  A suitable asymmetry is given
                 by~\cite{angle} 
\be 
O^b_{\lambda\lambda'} = 
 \frac{ N[ (\hat{\mathbf{q}}_{e} \hat{\mathbf{p}}_{t} 
\hat{\mathbf{p}}_{b})> 0]  -
N[( \hat{\mathbf{q}}_{e} \hat{\mathbf{p}}_{t}  \hat{\mathbf{p}}_{b} )<
                  0]}  {N[ ( \hat{\mathbf{q}}_{e} \hat{\mathbf{p}}_{t}
            \hat{\mathbf{p}}_{b}  ) > 0]  + N[ (  \hat{\mathbf{q}}_{e}
 \hat{\mathbf{p}}_{t}  \hat{\mathbf{p}}_{b} ) < 0] } \: , 
\ee 
where 
 $N[ (  \hat{\mathbf{q}}_{e} \hat{\mathbf{p}}_{t}) \hat{\mathbf{p}}_{b}
   ) > 0 (<0)$] are the number of $b$ quarks produced above/below  the
                                            production--plane  {\small
$\{\vec{\mathbf{q}}_{e},\vec{\mathbf{p}}_t\}$}  at a  given
                                      polarization $\lambda,\lambda'$.

        As in general $O_{\lambda \lambda'}^b$  gets also CP invariant
contributions from  absorptive parts in the SM amplitude, the truly CP
   violating  contribution will be singled out through the difference:
  \be 
 {\cal O}_b^{CP}(\lambda\lambda') =  O^{b}_{\lambda,\lambda'} - O^{\bar
 b}_{-\lambda',-\lambda} =  - \alpha_{b} \frac{3 \beta \pi \sqrt{s}}{2
                                    m_{t}}  \frac{\eRe D_{1}}{N_{tot}}
\label{OCPb} 
          \ee  
In the above equation  $O^{\bar{b}}$  refers to process
            (\ref{b}) when the $\bar t$ decays.  A non--zero value of
       (\ref{OCPb}) would imply CP violation in the  $t\bar{t}
                                  \gamma$ and/or $t\bar{t}Z$ vertices.

   For the above asymmetries we have obtained expressions in which all
            phase space integrations have been performed analytically.

\section{The process $e^+ e^- \rightarrow blX$}

\subsection{The differential cross section}  

    From (\ref{sigmal}) and the expressions for $\xi^t$ and $\xi^{\bar
          t}$  we can obtain  
 the differencial cross section $d\sigma^\pm$ of processes
(\ref{l}) in the c. m. system. If we keep in (\ref{sigmal}) 
only the components of $\xi^{t,\bar t}_{CP}$ that are normal 
to the production plane, i. e. the terms proportional 
to $D^t$ and $D^{\bar t}$ in eqs. (\ref{xit}) and 
(\ref{xitbar}), we shall obtain the dependence of $d\sigma^{\pm}$ 
on the triple product correlations of type (\ref{triple}).  In the
        c.m.system we have~\cite{we}: 
\be 
 d\sigma^{\pm}_{\lambda\lambda'} =
   \sigma_{SM}^{\pm}\left\{1\right.  &+& \frac{1}{1-\beta (\hat{{\bf
     p}}_t\hat{{\bf p}}_{l^\pm})} \left[\left(\hat{{\bf q}}_e\hat{{\bf
        p}}_t\hat{{\bf p}}_{l^\pm}\right)C^{\pm}_1 + \left(\hat{{\bf
  q}}_e\hat{{\bf p}}_t\hat{{\bf p}}_b\right)C^{\pm}_2\right.\nn 
 & +&\left(\hat{{\bf p}}_t\hat{{\bf p}}_{l^+} \hat{{\bf
      p}}_b\right)C^{\pm}_3 +\left.\left.\left(\hat{\bf q}_e\hat{\bf
     p}_{l^\pm}\hat{\bf p}_b\right)C^{\pm}_4\right]\right\}d\Omega_t
  d\Omega_b d\Omega_l \label{sigmatriple}  
\ee 
 where $\hat{\bf q}_e ,
         \hat{\bf p}_t$, etc. denote the corresponding unit 3-vectors.
      $\sigma_{SM}^{l^\pm}$ determines the expression for the SM cross
   sections of (\ref{l}): 
\be 
 \sigma_{SM}^{\pm}(\lambda ,\lambda' )
      =\sigma_0^{\pm}(\lambda ,\lambda' ) A_{SM}^{\pm}\,.
 \ee 
The expressions for $\sigma_0^\pm$ and $A_{SM}^\pm$ are 
given by (\ref{sigmal0}) and (\ref{ASMl}).  
 For the        functions  $C_i$ we obtain
\footnote{Note the  i in the definitions of
  $\dgs$ and $\dzs$ in (\ref{vertexgamma}) and (\ref{vertexZ}) 
that leads to the appearance of the real form factors in $C_i$  
in stead of the imaginary ones in \cite{we}.}: 
 \be 
 C^{\pm}_1
            &=&\mp\beta\left[\frac{{\cal D}^{t(\bar t)}({\it CP} )}{2}-
       \frac{m_t}{\sqrt s}E_{b(\bar b)} {\cal P}_-^{t(\bar t)}({\it SM})
    \frac{\Re e g_R^{CP}}{m_W}\right]\label{C1} \\
 C^{\pm}_2 &=&
  \mp\beta\frac{m_t}{\sqrt s} E_{b(\bar b)} {\cal P}_ -^{t(\bar t)}({\it
   SM})\frac{\Re e g_R^{CP}}{m_W}\label{C2} \\
  C^{\pm}_3 &=&\mp
      \beta\frac{m_t}{\sqrt s}E_{b(\bar b)}  {\cal P}_+^{t(\bar t)}({\it
        SM})\frac{\Re e g_R^{CP}}{m_W}\label{C3}\\
 C^{\pm}_4 &=&
  \pm\frac{m_t} {\sqrt s} E_{b(\bar b)} {\cal P}_-^{t(\bar t)}({\it SM})
        \frac{\Re e g_R^{CP}}{m_W}\label{C4}  
\ee 
where ${\cal P}_
  \pm^{t(\bar t)}$ and ${\cal D}_ \pm^{t(\bar t)}$,
introduced previously in (\ref{calPD}) are given by 
\be
&& {\cal P}_ +^{t(\bar
   t)}({\it SM}) =-\frac{4}{N^{t(\bar t)}_{\lambda\lambda'}} \left[\pm
          G_3+\beta\cos\theta_{t(\bar t)}\, (G_1-G_2)\right]\nn
&& {\cal P}_ -^{t(\bar t)}({\it SM})=
              \frac{4}{N^{t(\bar t)}_{\lambda\lambda'}}\left[G_1+G_2\pm
      \beta\cos\theta_{t(\bar t)}\, G_3\right]\nn 
&& {\cal D}^{t(\bar t)}({\it CP}) =\frac{8}{N^{t(\bar t)}_{\lambda\lambda'}}
      \left[D_1\pm\beta\cos\theta_{t(\bar t)}\, D_2\right]\,.\nonumber
\ee 
 The result of (\ref{C1}) - (\ref{C4}) can be easily
    uinderstood. The correlations $(\hat{\bf q}_l\hat{\bf p}_t\hat{\bf
   p}_l)$ and $(\hat{\bf q}_l\hat{\bf p}_t\hat{\bf p}_b)$ contain  the
 production and $t$-decay planes, and thus CP violation from both  the
      production and the decay may contribute to $C_1$ and $C_2$.  The
       triple products $(\hat{\bf p}_t\hat{\bf p}_l\hat{\bf p}_b)$ and
    $(\hat{\bf q}_l\hat{\bf p}_l\hat{\bf p}_b)$ contain only the decay
plane, and thus  CP violation only in the decay vertex may enter $C_3$
     and $C_4$.  From the explicit expressions for $C_i$,  one can see
           however that $C_2$ gets no contribution from the production
    process. This is a result of a  cancelation due to  the same $V-A$
    form of the $tbW$ and $l\nu W$ vertices. The terms proportional to
   $\Re e g_R^{CP}$ always enter multiplied by the SM-polarization and
   the kinematic factor $E_bm_t/\sqrt s$. Consequently $C_1$, the only
    term that contains $\Re ed^\gamma$, $\Re ed^Z$ is the dominat one.

    The $\cos\theta_l$-distribution of the decaying leptons, that will
depend on $\Im m d^{\gamma ,Z}$ was obtained in \cite{Rindani}. 
The analytic expression  
  is the same as that for the $\cos\theta_b$-distribution, 
eq.(\ref{angle}), but for the replacement
$\alpha_b \rightarrow \alpha_l$, $\Gamma_{t\to bW} \rightarrow
 \Gamma_{t\to bl\nu}$. This can be understood  having in mind the
       same $\gamma_\alpha (1-\gamma_5)$ form of the $tbW$ 
and $l\nu W$  vertices.

Analytic expressions for the energy distribution of the secondary
 leptons $l^+$ and $l^-$ in case of CP violation in the production
 process were obtained in \cite{{Chang},{Sehgal-2}}
 and later, including also CP violation in the decay 
in \cite{Grz-E}. Observables sensitive to CP violation in
 the production and the decay are discussed in \cite{Grz-E}.

\subsection{Triple product correlations}

With the set of triple products ${\cal T}_1 =  ({\bf q_ep_tp_{l^+}})$,
              ${\cal T}_2 =  ({\bf q_ep_tp_b})$,  ${\cal T}_3 =  ({\bf
       p_tp_{l^+}p_b})$, and ${\cal T}_4 =  ({\bf q_ep_{l^+}p_b})$, we
 define the  following observables for processes (\ref{l}): 
\be 
O_i^t&
      =& \frac{N[{\cal T}_i >0] - N[{\cal T}_i<0]} {N[{\cal T}_i >0] +
    N[{\cal T}_i<0]} 
\ee 
where $N[{\cal T}_i >0(<0)]$ is the number of
    events in which ${\cal T}_i>0(<0)$. For example a nonzero value of
   $O_1^t$ would mean that there is a difference between the number of
       events in which $l^+$ are above and bellow the production plane
                                              (${\bf q}_e,{\bf p}_t$).

 As $O_i^t$ are T-odd asymmetries, they get contributions from final
   state interactions, too. The truly CP violating observables are the
             differencies: 
\be 
 {\cal O}^{CP}_i = O_i^t - O_i^{\bar t}, \quad
  i=1,2,4\qquad {\rm and}
\qquad {\cal O}_3^{CP} = O_3^t + O_3^{\bar t}\,, 
\label{OCPl}
\ee 
where
     $O_i^t$ refer to process (\ref{l}) when the $t$-quark decays, and
   $O_i^{\bar t}$ refer to  process (\ref{l}) when $\bar t$ decays.

 As the considered triple products are not orthogonal, each observable
     ${\cal O}_i^{CP}$ will get in general contributions from all
 functions $C_k$,   $k=1,2,3,4$.  From the explicit expressions
(\ref{sigmatriple})-(\ref{C4})  one can show that $C_1$ enters the
  asymmetries ${\cal O}_1^{CP}, {\cal O}_2^{CP}$ and ${\cal O}_4^{CP}$,
 and being the
 dominat contribution it determines thier magnitude and sensitivity to
     CP violation in the production process. To ${\cal O}_3^{CP}$ 
the terms  $C_3$ and $C_4$ contribute, 
and this implies that it is sensitive to
 CP violation in the decay process only. Therefore, a nonzero value of
 ${\cal O}_1^{CP}, {\cal O}_2^{CP}$ and ${\cal O}_4^{CP}$
 will be an indication of CP
   violation in the production plane, and ${\cal O}_3^{CP}$ will measure CP
   violation in the decay. However, because of the suppression factors
  ($t$-polarization and  kinematics) ${\cal O}_3$ will be too small to
                                                 measure CP violation.
This model independent analysis was confirmed by our numerical results 
 performed in the MSSM~\cite{we}.

\section{CP violation in the $t$ decay vertex}

    The matrix elements for $t\rightarrow bW^+$ and $\bar t\rightarrow
     \bar b W^-$ in case of CP violation are: 
\be 
 M_t = {\bar u}(p_b)
     V_\alpha^t u(p_t) \epsilon^\alpha  (p_{W^+})\quad {\rm and} \quad
     M_{\bar t} = {\bar u}(p_{\bar b}) V_\alpha^{\bar t} u(p_{\bar t})
     \epsilon^\alpha(p_{W^-})\,,\label{Mtop} 
\ee 
where $V^{t,\bar t}$ are given by
          (\ref{t}) and (\ref{tbar}). 
 As shown in Sect. 2, eqs. (ref{sigmab}) and (\ref{sigmal}) 
imply  that the  energy and angular
     distributions of the $b$ quarks and the leptons, 
 including the possible triple product correlations are actually
         sensitive to CP violation in the production process only.  CP
     violation in the decay $t$-quark vertex  leads to a nonzero value
between the partial decas widths of $t\to bW$ and $\bar t\to \bar b W$
--      the asymmetry $A_{CP}$,  eq.(\ref{CPdecay}).
 From the interference of the tree level amplitude and 
the loop corrections containingthe terms $f_L$ and $g_R$,  we obtain
 \cite{ECMF3}: 
\be 
A_{CP} =2\left[ \Im m\, f_L^{CP} + \Im m\, g_R^{CP}
      \frac{m_t(m_t^2 -m_W^2)}{m_W(m_t^2 + 2m_W^2)}\right]\,.
\label{ACP}  
\ee 
 In  processes (\ref{b}) and (\ref{l}) this quantity can be measured by
the asymmetries $\Delta^b $ and $\Delta^l$ respectively: 
\be 
\Delta^{b} = \frac{N_{tot}^b(\lambda ,\lambda') -
            N^{\bar b}_{tot}(-\lambda' ,-\lambda)}
{N_{tot}^b(\lambda ,\lambda') +
            N^{\bar b}_{tot}(-\lambda' ,-\lambda)}
= A_{CP},\nonumber
\ee
\be
 \Delta^{l} =
\frac{N_{tot}^+(\lambda ,\lambda') -
            N^-_{tot}(-\lambda' ,-\lambda)}
{N_{tot}^+(\lambda ,\lambda') +
            N^-_{tot}(-\lambda' ,-\lambda)}= A_{CP}.        
\label{deltaCP}  
\ee 
Here
  $N^{b(\bar b)}_{tot}$ and $N^{\pm}_{tot}$ are the total number of 
 $b(\bar b)$
            quarks produced in (\ref{b}), and of $l^{\pm}$ produced in
      (\ref{l}). $\Delta^b$ and $\Delta^l$ measure CP violation in the
          $t$-decay vertex only, irrespectively of CP violation in the
production process.  Note that if $t\to bW$ is  the only decay mode of
 the $t$-quarks,  as it is actually in SM, $\Delta^b$ would be zero by
                                                      the CPT theorem.

\section{Numerical results in MSSM}

      Up to now our expressions for the asymmetries were general and model
    independent. Here we shall give numerical results for the CP violating
          asymmetries in the Minimal Standard Supersymmetric Model (MSSM).

 In the SM CP violation appears through the phase of the CKM mixing matrix
 only if the three genetarions of quarks mix. 
 This contribution is small, restricted
    by the unitarity condition on the mixing matrix. Further, again due to
    unitarity the dipole moment form factors $d^\gamma $ and $d^Z$ are at
                least two-loop effect and hence of academic interest only.
The contribution of the self-energy loop in SM to the CP violating 
vertex was also shown to be extremely small \cite{Grz-decay}.

         In the Lagrangian of the MSSM additional complex couplings are 
     introduced \cite{Dugan} that lead to CP violation within
 one generation only. This
   leads to CP violation at one loop,   free of the unitariry suppression.
  As the masses of the SUSY particles  are not expected to be much heavier
    than the mass of the top quark, the radiative corrections through
      which the CP violating form factors are induced will not be strongly
                    suppressed by the masses of the particles in the loop.

    There are two physical complex couplings in the MSSM Lagrangian -- the
 parameter $\mu$ in front of the Higgsino mass term, and the
 dimensionless parameter $A_f$  in  the soft SUSY breaking piece of the
 Lagrangian. The magnitude of the CP violating form factors depend strongly
         on the phases of these parameters. The parameter $A_f$ depends on
  flavour. Measurements of the electric dipole moment (EDM) of the neutron
put constraints on some of these phases \cite{Osh}. This is the so called 
supersymmetric  fine-tuning CP problem -- 
 usually one concludes that either
the phases involved in the EDM of the neutron are very small or the masses
       of the first generation of the squarks are in the TeV region. 
A complete analysis of the constraints on the SUSY parameters from
 measurements of the EDM's of the neutron and the electron was done
 in~\cite{Nath}. Using
              supergravity with grand unification (GUT) there are attempts
 \cite{Garisto}  to constrain also the phases of $A_t$ and $A_b$ that
    enter the CP violating form factors of the top quark. For our numerical
    analysis we shall not make any additional assumptions about GUT except
  unification of the gauge couplings, i.e. we do not assume unification of
the scalar mass parameters and the parameters $A_f$. As the mechanism of 
SUSY breaking is not known, an unambigous decision about CP violation in
 SUSY will be provided by experiment.   


\subsection{The asymmetries due to $d^\gamma$ and $d^Z$}

In order to estimate the observables sensitive to CP violation in
 the production process we have used the results for the electro weak 
dipole moment form factors $d^\gamma$ and $d^Z$ as obtained in 
\cite{dipole}, where  a complete analysis was performed  with gluino, 
charginos and neutralinos exchanged in the loops. 

 There are two types of observables -- sensitive to 
$\Re ed^{\gamma ,Z}$ and to $\Im md^{\gamma ,Z}$. As an illustration,  
on Fig.1 and Fig.2  the values of 
${\cal A}^{FB}_{\lambda \lambda'}$, eq (\ref{AFB}), 
 and the values of 
${\cal O}_{b}^{CP}$, eq. (\ref{OCPb}), 
 as functions of the c.m. energy 
$\sqrt s$ are shown. The asymmetry ${\cal A}^{FB}_{\lambda \lambda'}$
 is determined by  
the dependence on $\sqrt s$ of $\Im md^{\gamma ,Z}$, and
 the asymmetry ${\cal O}_{b}^{CP}$ -- by  $\Re ed^{\gamma ,Z}$. 
The figures are presented for the following values of the SUSY  parameters: 
$ M= 230$ GeV, $\vert\mu\vert = 250$ GeV, $m_{\tilde{t}_1}=150$ GeV,
$m_{\tilde{t}_2}=400$ GeV, 
 $ m_{\tilde{b}_1}=270$ GeV and 
$m_{\tilde{b}_2}=280$ GeV. The following GUT relations 
between the gaugino mass parameters have been assumed:    
 $m_{\tilde{g}} = (\alpha_s/\alpha_2 )M \approx  3M$ and 
$M'= (5/3) \tan^2\Theta_W M$.

Clearly seen are the spikes in the $\sqrt s$ behaviour of 
  ${\cal A}^{FB}_{\lambda \lambda'}$ and 
 ${\cal O}^{b}_{CP}$. They  
are  due to the  threasholds of the intermediate particles in the loop and 
  are  already present in the dipole moment form factors 
$\Im m d^{\gamma ,Z}$ and $\Re ed^{\gamma ,Z}$ as discussed in detail 
in \cite{dipole}. Their position is determined by the
 spectra of the particles in the loop, their magnitude --
 by the strength of the couplings. For example, the spikes at 
$\sqrt s= 400 GeV $ and $\sqrt s= 590 GeV$ are due to the
 threasholds of 
$\tilde{\chi}_1^+\tilde{\chi}_1^-$ production with 
$m_{\tilde{\chi^+_1}} = 200 GeV$ and of 
$\tilde{\chi}_2^+\tilde{\chi}_2^-$ production with 
$m_{\tilde{\chi}^+_2} = 295$ GeV, respectively. 
The asymmetries depend strongly on the polarization of 
the electron and positron beams $\lambda ,\lambda'$. 
The figures show the asymmetries 
 for different beam polarizations, taking
 $\lambda = -\lambda' = (-0.8,0,0.8)$. 
Notice the  strong dependence  on the polarization of the electrons. For  
 $\sqrt s < 700 GeV$ the asymmetreis are much bigger  if 
the electrons are left handed. This is more clearly pronounced for 
${\cal O}^{b}_{CP}$. In general the asymmetries are of the 
order of $10^{-3}$. 

\subsection{The asymmetries due to $f_L^{CP}$ and $g_R^{CP}$}

The asymmetry $A_{CP}$ is determined by $\Im mf_L^{CP}$ and $\Im mg_R^{CP}$
 - eq. (\ref{ACP}). In order that $\Im mf_L^{CP}$ and $\Im mg_R^{CP}$
 are  different form zero, we need at the same time, CP 
violating complex couplings in the
       Lagragian, provided here by the MSSM. 
and non vanishing absoptive parts in the loop integrals.
   Since only the absorptive
parts of the loop SUSY amplitude enter  $\Im mf_L^{CP}$ and $\Im mg_R^{CP}$, 
 the main contribution
         would come from diagrams in which one of the on -- shell loop
  particles is the lightest SUSY particle -- the neutralino
 $\tilde{\chi}^0_1$. There are two such diagrams: with
                ($\tilde{\chi}^0_1-\tilde{t}_1-\tilde{\chi}^+_i$) 
and with 
($\tilde{t}_1-\tilde{\chi}^0_1-\tilde{b}_L$) in the loop ($\tilde{t}_n$
  are the massive scalar-top states, $\tilde{\chi}^+_i$ are  the
 chargino states,
    the mass of the $b$ -- quark has been neglected). 
The present 
 experimental bounds on the gluino and the scalar top masses 
forbids kinematically the diagram with 
($\tilde{t}_L-\tilde{g}-\tilde{b}_L$) in the
 loop that  could lead to a big   contribution \cite{Grz-decay}.   
 Full expression of the 
            contribution from the different diagrams was  obtained
         in~\cite{ECMF3}, where also numerical integration was
         performed. Detail analysis of the dependence on the SUSY mass
parameters was carried out in~\cite{Oshimo}.

 The full expression is rather combursome, however a  rather simple
    expression, that gives the order of magnitude of the effect can be
                                     obtained if only the diagram with
($\tilde{t}_1-\tilde{\chi}^0_1-\tilde{b}_L$) is considered. 
For a very light neutralino, if we  neglect 
 the mixing between the gaugino components  $\tilde {W}_3$ and $\tilde
  {B}$ as compared to that between the gaugino and Higgsino components
     $\tilde {W}_3 (\tilde{B})$ and $\tilde{H}^0$, as suggested by the
       minimal supergravity models~\cite{Kane} and 
parametrize the possible imaginary couplings by introducing a 
single CP violating phase  $\sin\delta_{CP}$  we obtain~\cite{ECMF3}:
\be
A_{CP} \simeq&&-\frac{\alpha_{em}}{\sin^2\Theta_W}
 \frac{\sqrt{2}}{4 \sin \beta}
\frac{m_t^2}{m_t^2 +
 2 M_W^2} 
\frac{m_{\tilde{\chi}^0_1}}{M_W}\times\nn
&&\qquad\qquad\times \ln \left\vert 1 - \frac{(m_t^2 - M_W^2)
(m_t^2 -m_{{\tilde t}_1}}{m_t^2 m_{\tilde{b}_L}^2}
 \right\vert \sin \delta_{CP} \, .
\ee

 For maximal
$CP$ violation ($\sin \delta_{CP} = 1$),
  $ m_{\tilde{t}_L} =150$
GeV and $m_{\tilde {\chi}^0}=20$ GeV (near the exprimental bound), we have:
\be
A_{CP} \simeq 0.059 \times \frac{\alpha_{em}}{\sin^2\Theta_W},\,{\rm for}\,\,
 m_{\tilde{b}_L}=200 GeV
\ee
\be
A_{CP}\simeq 0.026 \times \frac{\alpha_{em}}{\sin^2x\Theta_W} \,
\,{\rm for}\,\,
 m_{\tilde{b}_L}=300 GeV.
\ee
which is an asymmetry of the order of $10^{-3}$.
This value is obtained also taking into account the constraints from the 
EDM's of the neutron and the electron~\cite{Oshimo}.

\section{Conclusions}

 CP violation in the $\gamma t \bar t$
 and $Zt\bar t$ vertices in  the production 
process $e^+ e^- \to t \bar t$, and CP violation in the $tbW$ vertex 
 in the $t$-decays $t\to bW$ or 
$t\to bW \to bl^+\nu$ have been assumed. 
Studying the single  $b$-quark and lepton distributions 
 we have defined asymmetries that can disentangle
 CP violation in the production from CP violation in the decay.  
The angular and energy asymmetries are actually sensitive 
to CP violaion
 in the production process only, CP violation in the decay being
 suppressed  by the amount of the SM top 
quark polarization, for the secondary leptons it is suppressed
 also kinematically. 
 Appropriate angular and energy 
asymmetries  that determine  independently the real 
and imaginary parts of the dipole moment form factors  
$d^\gamma (s)$ and $d^Z(s)$ are defined. 
 CP violation in the decay can be measured 
 through the difference between the total number of 
$b$ and $\bar b$ quarks or $l^+$ and $l^-$ from the decay of the $t$ and 
$\bar t$ quarks.  
 Particular attention is paid 
to the polarization of the top quark.

Analytic expressions for the  considered distributions and 
asymmetries are obtained.  
These expressions are general and independent on the definite model
 of CP violation.
 In the formula  that involve only the $b$ quarks all phase space
 integrations have been carried out analytically. 
 A  numerical analysis of the asymmetries 
have been performed in the Minimal Standard Supersymmetric Model 
with complex phases. In this model the effects turn out to be of the
 order of $10^{-3}$. With the planned luminosities for 
the $e^+e^-$ linear collider this is on the borderline of detectability.
However other models of CP violation, for example the two-Higgs doublet 
model, can give much higher asymmetries \cite{2Higgs}. 
Observation of the asymmetries discussed above will
 be interesting as it would be a definite signal of physics beyond the SM. 

\section*{Acknowledgements}
\addcontentsline{toc}{section}{Acknowledgements}

I am very thankful to Alfred Bartl, Marco Fabbrichesi,
 Thomas Gajdosik and Walter Majerotto
 with whom the above results have been obtained and 
 to Thomas Gajdosik for his ready help in preparing the pictures. 
This research is supported by the Bulgarian Science Foundation,
 Grant Ph-510.



\section*{Figure Captions}
\addcontentsline{toc}{section}{Figure Captions}

\begin{enumerate}

\item FIG. 1 The asymmetry ${\cal A}_{\lambda \lambda'}^{FB}$
 that measures $\Im md^{\gamma ,Z}$, 
 eq. (\ref{AFB}) as a function of $\sqrt s$ (GeV) for different
 beam polarizations: $\lambda '=-\lambda$; 
$\lambda$ = -0.8 (dashed line),0  (full line) and  0.8  
 (dotted line).  

\item 
FIG. 2 The asymmetry ${\cal O}^{CP}_b (\lambda \lambda')$  
 that measures $\Re ed^{\gamma ,Z}$,
 eq. (\ref{OCPb}) as a function of $\sqrt s$ (GeV) for different
 beam polarizations: $\lambda '=-\lambda$; 
$\lambda$ = -0.8  (dashed line), 0  (full line) and  0.8  
 (dotted line).  
\end{enumerate}

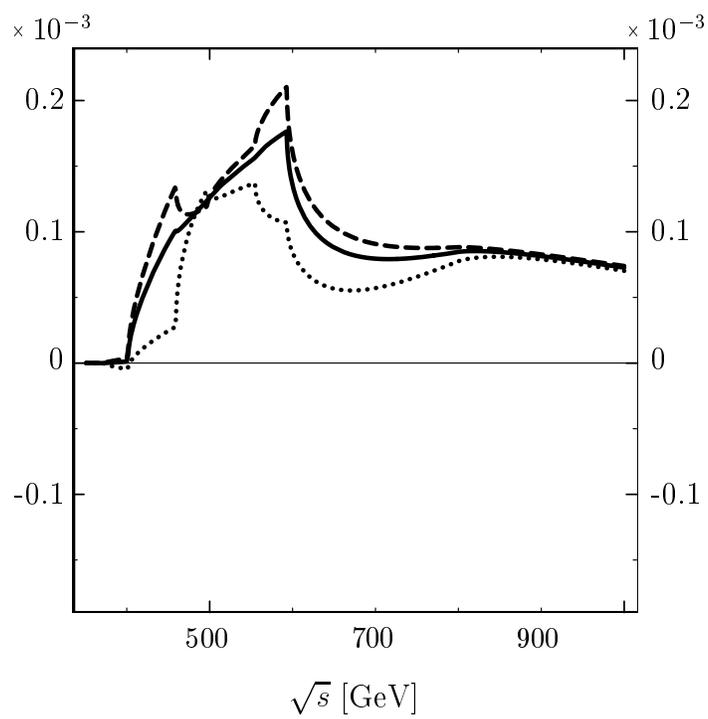
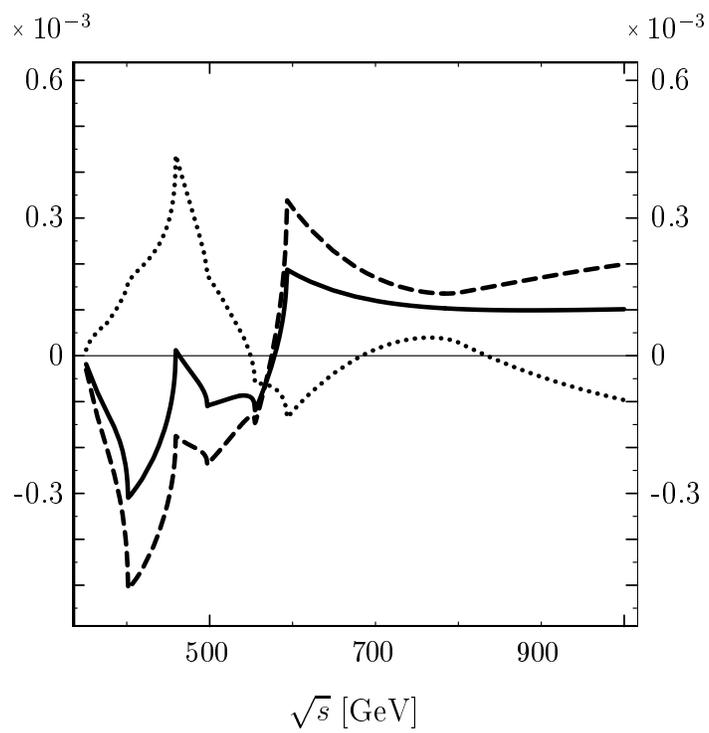
\end{document}